\documentclass[trackchanges]{aastex7}

\makeatletter
\def\endfigure{\end@float}
\makeatother

\usepackage{lipsum}
\usepackage{mwe}

\usepackage{lipsum,graphicx}
\usepackage{newtxtext,newtxmath}
\usepackage[T1]{fontenc}
\usepackage{ae,aecompl}\usepackage{fix-cm}
\usepackage{adjustbox}
\usepackage{amsfonts}
\usepackage{amsmath}
\usepackage{graphicx}	
\usepackage{tabularx}
\usepackage{verbatim}
\usepackage{xspace}
\usepackage{adjustbox}
\usepackage{rotating, lipsum, babel}
\usepackage{MnSymbol}

\usepackage{subfig}

\newcommand{\pc}{\,\mathrm{pc}}
\newcommand{\Msun}{\,\mathrm{M}_{\odot}}
\newcommand{\kpc}{\,\mathrm{kpc}}
\newcommand{\Gyr}{\,\mathrm{Gyr}}

\newcommand{\kms}{\,\mathrm{km\,s}^{-1}}

\newcommand{\Mv}{M_\mathrm{v}}
\newcommand{\Nbody}{$N$-body\xspace}
\newcommand{\Rh}{R_\mathrm{h}}

\newcommand{\secref}[1]{Section~\ref{#1}}
\newcommand{\figref}[1]{Figure~\ref{#1}}
\newcommand{\tabref}[1]{Table~\ref{#1}}

\graphicspath{{Figures/}}

\begin{document}

\title{Dark Star Clusters or Ultra-Faint Dwarf galaxies? Revisiting UMa3/U1}

\author{Ali Rostami-Shirazi}
\affiliation{Department of Physics, Institute for Advanced Studies in Basic Sciences (IASBS), 444 Prof. Sobouti Blvd., Zanjan 45137-66731, Iran}
\email[show]{a.rostami@iasbs.ac.ir}

\author{Hosein Haghi}
\affiliation{Department of Physics, Institute for Advanced Studies in Basic Sciences (IASBS), 444 Prof. Sobouti Blvd., Zanjan 45137-66731, Iran}
\affiliation{Helmholtz-Institut f\"ur Strahlen-und Kernphysik (HISKP), Universit\"at Bonn, Nussallee 14-16, D-53115 Bonn, Germany}
\email[show]{Haghi@iasbs.ac.ir}

\author{Akram Hasani Zonoozi}
\affiliation{Department of Physics, Institute for Advanced Studies in Basic Sciences (IASBS), 444 Prof. Sobouti Blvd., Zanjan 45137-66731, Iran}
\affiliation{Helmholtz-Institut f\"ur Strahlen-und Kernphysik (HISKP), Universit\"at Bonn, Nussallee 14-16, D-53115 Bonn, Germany}
\email{}

\author{Pavel Kroupa}
\affiliation{Helmholtz-Institut f\"ur Strahlen-und Kernphysik (HISKP), Universit\"at Bonn, Nussallee 14-16, D-53115 Bonn, Germany}
\affiliation{Charles University in Prague, Faculty of Mathematics and Physics, Astronomical Institute, V Hole\v{s}ovi\v{c}k\'ach 2, CZ-180 00 Praha 8, Czech Republic}
\email{}

\begin{abstract}

Owing to sparse spectroscopic observations, the classification of faint satellites as either dark matter–dominated dwarf galaxies or self-gravitating star clusters remains unresolved. The recently discovered Ursa Major III/UNIONS 1 (UMa3/U1) object, with its measured velocity dispersion, provides a rare observational anchor in this regime. Despite its cluster-like compactness, its inferred dynamical mass-to-light ratio ($M_\mathrm{Dyn}/L$) suggests a dark matter–dominated nature, prompting interpretations of UMa3/U1 as a microgalaxy, though current measurements remain inconclusive. Thousand-level $M_\mathrm{Dyn}/L$ values are not unique to galaxies; self-gravitating dark star clusters (DSCs) can reach comparable levels via energy injection driven by a centrally segregated black hole subsystem (BHSub), which accelerates the evaporation of luminous stars and leads to a super-virial appearance with elevated velocity dispersion. To assess whether UMa3/U1 is a DSC, we conducted direct \Nbody simulations and identified a model that successfully reproduces both its compact structure and elevated $M_\mathrm{Dyn}/L$, supporting a self-gravitating cluster origin. We find the cluster entered the DSC phase around $4 \Gyr$ ago, with its luminous stars expected to be depleted within the next $1\Gyr$, followed by the gradual disruption of the central BHSub over the subsequent Gyr.We broaden our analysis by mapping DSC evolutionary tracks in the size versus total luminosity ($L$) and $M_\mathrm{Dyn}/L$–$L$ spaces, showing that DSCs occupy a region overlapping with faint, ambiguous satellites. In the $M_\mathrm{Dyn}/L$–$L$ diagram, DSCs trace a transitional channel bridging globular clusters and dwarf galaxies as they rise from $M_\mathrm{Dyn}/L \approx 2$ to $10^4 \ \mathrm{M_{\odot}/L_{\odot}}$.

\end{abstract}


\section{Introduction}\label{sec:intro}

In the hierarchy of stellar systems, dwarf galaxies (dGs) and globular clusters (GCs) stand out following galaxies in scale, exhibiting fundamental contrasts in their structure and composition. While both act as satellites of galaxies, GCs themselves may also have served as satellites of dGs (see e.g. \citealt{massari2019,Kruijssen2020,Rostami2022,Shirazi2023}). GCs are collisional systems that dynamically relax within a period shorter than a Hubble time, whereas dGs, being collisionless, evolve over timescales longer than a Hubble time \citep{Kroupa1998, Binney2008,Forbes2011,spitzer2014}. GCs typically display homogeneity in their stellar population's heavy-element abundances, whereas dGs, contrastingly, contain stars with significant metallicity dispersions \citep{Willman2012,Leaman2012,Kirby2013,Bastian2018, WangKroupa2020}. Although the internal kinematics of GCs adhere to a combination of baryonic matter and Newtonian dynamics \citep{Illingworth1976,Peterson1986}, the mass inferred from the luminosity of dGs is several orders of magnitude lower than estimates derived from kinematic considerations, yielding dynamical mass-to-light ratios $M_\mathrm{Dyn}/L \approx 10^1-10^4 \ \mathrm{M_{\odot}/L_{\odot}}$ \citep{Mateo1998,McConnachie2012,Simon2019}, whereby some of the dGs can be long-lived non-virial remnants of satellite galaxies lacking dark matter \citep{Kroupa1997, Casas2012}. For comparison, GCs typically exhibit $M_\mathrm{Dyn}/L\approx 2 \ \mathrm{M_{\odot}/L_{\odot}}$ \citep{McLaughlin2005,Strader2009,Baumgardt2020}.

Moreover, their distribution in the size versus total luminosity ($L$) plane reveals a bimodal accumulation. Despite similar $L$, dGs are significantly more diffuse than GCs \citep{Harris1996,McConnachie2012}. Over the past two decades, the discovery of faint stellar systems with ambiguous natures has bridged the gap between GCs and dGs in the size-luminosity plane, blurring the distinction at lower luminosities \citep{Willman2005,Koposov2007,Fadely2011,Munoz2012,Balbinot2013,Koposov2015,Conn2018,Torrealba2019,Mau2020,Cerny2023ref,Cerny2023Delve,Cerny2023,Smith2024,Simon2024}. Faint ambiguous satellites populate a region with projected half-light radii $\Rh\lesssim20\pc$ and total luminosities $L\lesssim10^3 \ \mathrm{L_{\odot}}$ \citep{Mau2020,Cerny2023ref,Smith2024}, where extrapolated sequences of both GCs and dGs would converge at fainter magnitudes. Spectroscopic velocity dispersion measurements are essential for distinguishing whether these tiny satellites are dark-matter-dominated dGs or self-gravitating GCs, although the limited spectroscopic sample size and unresolved binary stars challenge interpretation.

Ursa Major III/UNIONS 1 (hereafter UMa3/U1) is the faintest known Milky Way satellite, with a projected half-light radius of $\Rh=3\pm1\pc$ and a stellar mass of $M_{*}=16^{+6}_{-5} \Msun$, corresponding to an absolute $V$-band magnitude of $\Mv=+2.2^{+0.4}_{-0.3}$ mag, residing in the ambiguous region of the size-luminosity plane \citep{Smith2024}. It hosts an ancient ($\tau \gtrsim 11\Gyr$) and metal-poor ($\mathrm{[Fe/H]} \approx -2.2$) stellar population at a heliocentric distance of $\approx10 \kpc$ \citep{Smith2024}. UMa3/U1 is among the rare, faint, and ambiguous Milky Way satellites with follow-up spectroscopy, providing valuable insight into the nature of such systems. However, the potential contribution of binary stars to its velocity dispersion remains uncertain. Using a sample of 11 probable members, \citet{Smith2024} derived a line-of-sight velocity dispersion of $\sigma_\mathrm{los}=3.7^{+1.4}_{-1.0}\kms$. Excluding the furthest outlier in velocity drops the estimate to $1.9^{+1.4}_{-1.1}\kms$, implying a dynamical mass-to-light ratio of $M_\mathrm{Dyn}/L_{1/2}=1900^{+4400}_{-1600} \ \mathrm{M_{\odot}/L_{\odot}}$ within the half-light radius.

Assuming the observed size and stellar mass of UMa3/U1, \citet{Errani2024} used \Nbody simulations to show that a self-gravitating system in equilibrium would exhibit a velocity dispersion of only $\approx 50 \,\mathrm{m\ s}^{-1}$ and be disrupted by the Galactic tidal field within $\approx 0.4 \Gyr$ (approximately one orbital period) making its detection a finely-tuned event in time. In contrast, models of dGs embedded in cuspy dark matter halos account for UMa3/U1's long-term survival and reproduce its elevated velocity dispersion of 1–4$\kms$. As a result, \citet{Errani2024} argued the notion that UMa3/U1 is a dark-matter-dominated microgalaxy (see also \citealt{Micro}) is particularly interesting in view of the results based on the Chandrasekhar dynamical friction test for the existence of cold or warm dark matter particle halos based on the orbital motions of the Large and Small Magellanic Clouds (LMC, SMC, respectively) about the Milky Way \citep{Massana2022, OehmKroupa2024}  as well as on the results concerning galactic bars \citep{Roshan2021}.

Dark star clusters (DSCs) can also exhibit extremely high $M_\mathrm{Dyn}/L$, similar to those seen in some galaxies \citep{banerjee2011,Taylor2015,DSC,Ghasemi2024}. Retention of a substantial black hole (BH) population within the cluster, facilitated by low natal kicks, triggers the assembly of a centrally segregated BH subsystem (BHSub) due to the Spitzer instability \citep{spitzer2014}, which injects energy into the cluster through frequent BH interactions, thereby accelerating the evaporation rate of luminous stars \citep{Mackey2007,Mackey2008,Banerjee2010,Breen2013,Banerjee2017,Rostami2024-metallicity,Rostami2024-Remnant}. The BHSub-generated energy contributes to tidal stripping, and collectively, the BHs control the evaporation time of luminous stars. The DSC phase emerges when the BHSub's self-depletion timescale surpasses the evaporation timescale of luminous stars. During this phase, luminous stars appear observationally super-virial, as their velocity dispersion is enhanced by the unseen BHSub, leading to a high $M_\mathrm{Dyn}/L$ \citep{banerjee2011,DSC,Wu2024}.

In this paper, we use collisional \Nbody simulations to investigate whether an evolved cluster in the DSC phase can reproduce UMa3/U1’s observational features and elucidate its nature. Furthermore, we assess the broader viability of DSCs as a formation scenario for faint ambiguous satellites. The paper is organized as follows: \secref{sec:sim} outlines the initial setup of the \Nbody simulations. The main results are presented in \secref{sec:result}, followed by a discussion and conclusion in \secref{sec:conclu}.

\section{SIMULATIONS}\label{sec:sim}

To investigate whether self-gravitating star clusters that transition to the DSC phase could serve as plausible progenitors of UMa3/U1, we have conducted a suite of simulations using the collisional \Nbody code ‘\textsc{NBODY7}’ \citep{Aarseth2012}. The state-of-the-art \textsc{NBODY7} code incorporates many physical phenomena, including single and binary stellar evolution from the zero-age main sequence to their final stages \citep{Hurley2000,Belczynski2008,Belczynski2010}, as well as strong gravitational encounters involving binary stars \citep{mikkola1999}.

The initial positions and velocities of the simulated cluster's stars follow a Plummer phase-space distribution function \citep{plummer1911, Kroupa2008}, maintaining virial equilibrium. The initial stellar masses are distributed over a mass range of $0.07$–$150\Msun$, following a three-segment power-law initial mass function (IMF):
\begin{equation}\label{eq:top_heavy_imf}
    \xi(m) \propto m^{-\alpha}: 
	\begin{cases}
		\alpha_1 = 1.3 & 0.07\leq\frac{m}{\Msun}<0.5    \\
		\alpha_2 = 2.3 & 0.50\leq\frac{m}{\Msun}<1.0 \\
		\alpha_3 & 1.00\leq\frac{m}{\Msun}<150
	\end{cases}\\
\end{equation}
where $\alpha_3=2.3$ corresponds to the canonical IMF \citep{kroupa2001} and a top-heavy IMF has $\alpha_3< 2.3$ \citep{MarksMichael2012,Kroupa2013}. The adopted stellar metallicity is [Fe/H]$=-2.2$ \citep{Smith2024}.

The modeled clusters' orbits align with that of UMa3/U1. Taking the observed position (equatorial coordinates and heliocentric distance) and velocity (proper motion and line-of-sight velocity) of UMa3/U1 at face value \citep{Smith2024}, we traced its orbit backward in time to set the initial conditions of the models. The clusters are embedded within a static Galactic potential, consisting of three components: a central bulge, a Miyamoto–Nagai disc \citep{Miyamoto1975}, and a phantom logarithmic dark matter halo that is scaled to yield a circular velocity of 220$\kms$ at the solar radius. The bulge is modeled as a central point mass with a mass of $1.5\times10^{10}\Msun$, while the disc is characterized by a scale length of $4\kpc$ and a height of $0.5\kpc$ \citep{Read2006}, with a mass of $5\times10^{10}\Msun$ \citep{Xue2008}.

We set the initial mass ($M_\mathrm{i}$), three-dimensional half-mass radius ($r_\mathrm{h,i}$), and natal kick received by BHs in the simulated clusters to study their progression into the DSC phase. Following the definition by \citet{banerjee2011}, the onset of this phase occurs when observable luminous stars including nuclear-burning stars and white dwarfs within twice the tidal radius\footnote{The tidal (Jacobi) radius is estimated as \( r_\mathrm{t} = R_\mathrm{G} \left( \frac{M_{\mathrm{cl}}}{3M_{\mathrm{G}}} \right)^{1/3} \), where \( R_\mathrm{G} \) is the Galactocentric distance of the cluster, \( M_{\mathrm{cl}} \) is the bound cluster mass, and \( M_{\mathrm{G}} \) is the Galactic mass enclosed within \( R_\mathrm{G} \).} exhibit significant super-viriality, characterized by a virial coefficient greater than 1 (i.e., $Q_* \equiv -KE/PE>1$, where KE and PE are the kinetic and potential energy of the luminous stars that are in the cluster as calculated from all stars within a distance smaller than $r_\mathrm{t}$ from the cluster centre). Following \citet{DSC}, we define the scaled DSC lifetime as:
\begin{equation}
\widetilde{\tau}_\mathrm{DSC} = \frac{\tau_\mathrm{DSC}}{\tau_\mathrm{Diss}},
\end{equation}
where $\tau_\mathrm{DSC}$ denotes the duration for which the cluster remains in the DSC phase, and $\tau_\mathrm{Diss}$ is the total cluster lifetime (the time at which only ten stars remain gravitationally bound \footnote{A star is considered gravitationally bound to a system if its total energy, comprising its kinetic energy (relative to the center-of-mass velocity of the cluster or subsystem) and its gravitational potential energy (arising from all other stars in the system), is negative.} to the cluster). The value of $\widetilde{\mathit{\tau}}_\mathrm{DSC}$ determines how much of the cluster's lifetime is in the DSC phase. To maximize the chance of achieving the DSC phase, we assume near-complete retention of BHs formed in our models. BH natal kicks remain poorly constrained from both theoretical and observational perspectives. A widely used approach \citep{Belczynski2008, Fryer2012} adopts neutron star-like kicks drawn from a Maxwellian distribution, which are reduced by fallback that damps the ejected momentum. High-mass BHs formed via direct collapse—especially in metal-poor environments—thus receive negligible or no kicks. For our adopted metallicity, approximately $75~\%$ of the total BH mass is expected to remain kick-free \citep{Banerjee2020,Gieles2021}, supporting our assumption of high BH retention in line with fallback-regulated models. An overview of the performed simulations is given in \tabref{tab:initial_conditions}.

\begin{deluxetable*}{cccccccc}
\tablecaption{Initial conditions and final-stage outcomes of simulated clusters designed to evolve into the DSC phase.\label{tab:initial_conditions}}
\tablehead{
\colhead{Model} & 
\colhead{$M_\mathrm{i}$} & 
\colhead{$r_\mathrm{h,i}$} & 
\colhead{$\alpha_3$} & 
\colhead{$R_\mathrm{h}$} & 
\colhead{$M_\mathrm{Dyn}/L_{1/2}$} & 
\colhead{$\tau$} &
\colhead{$\widetilde{\mathit{\tau}}_\mathrm{DSC}$} \\
\colhead{} & 
\colhead{($10^5 \ \mathrm{M_{\odot}}$)} & 
\colhead{(pc)} & 
\colhead{} & 
\colhead{(pc)} & 
\colhead{($10^3 \ \mathrm{M_{\odot}/L_{\odot}}$)} & 
\colhead{(Gyr)} &
\colhead{}
}
\startdata
M1 & 0.2 & 5  & 2.3 & 1.6  & 0.009 & 6.4 & 0.23\\
M2 & 0.3 & 7  & 2.3 & 9.5  & 0.024 & 7.4 & 0.29\\
M3 & 0.5 & 10 & 2.3 & 11.7 & 0.065 & 8.0 & 0.30\\
M4 & 0.8 & 10 & 2.3 & 14.4 & 0.28  & 9.3 & 0.46\\
M5 & 1.0 & 8  & 2.3 & 3.5  & 1.00  & 13.4 & 0.35\\
M6 & 1.0 & 12 & 2.3 & 6.7  & 0.33  & 11.2 & 0.44\\
M7 & 1.0 & 1  & 1.7 & 8.7  & 15.0  & 6.0 & 1.00\\
M8 & 1.0 & 2  & 1.7 & 11.3 & 12.0  & 5.6 & 0.99\\
\hline 
UMa3/U1 & \nodata & \nodata & \nodata & $3^{+1}_{-1}$ & $1.9^{+4.4}_{-1.6}$ & $>11$ & \nodata \\
\enddata
\tablecomments{Columns: (1) model designation; (2) initial stellar mass ($M_\mathrm{i}$); (3) initial 3D half-mass radius ($r_\mathrm{h,i}$); (4) slope of the IMF’s high-mass end ($\alpha_3$); (5) projected half-light radius ($\Rh$); (6) dynamical mass-to-light ratio within the half-light radius ($M_\mathrm{Dyn}/L_{1/2}$); (7) cluster age ($\tau$); and (8) scaled DSC lifetime ($\widetilde{\tau}_\mathrm{DSC}$). Final quantities (5–7) are evaluated when the bound stellar mass drops to $M* = 16~\Msun$, with (5) and (6) averaged over nearby snapshots to smooth out fluctuations.}
\end{deluxetable*}

\section{Results}\label{sec:result}

\begin{figure*}
    \centering
    \includegraphics[width=1\linewidth]{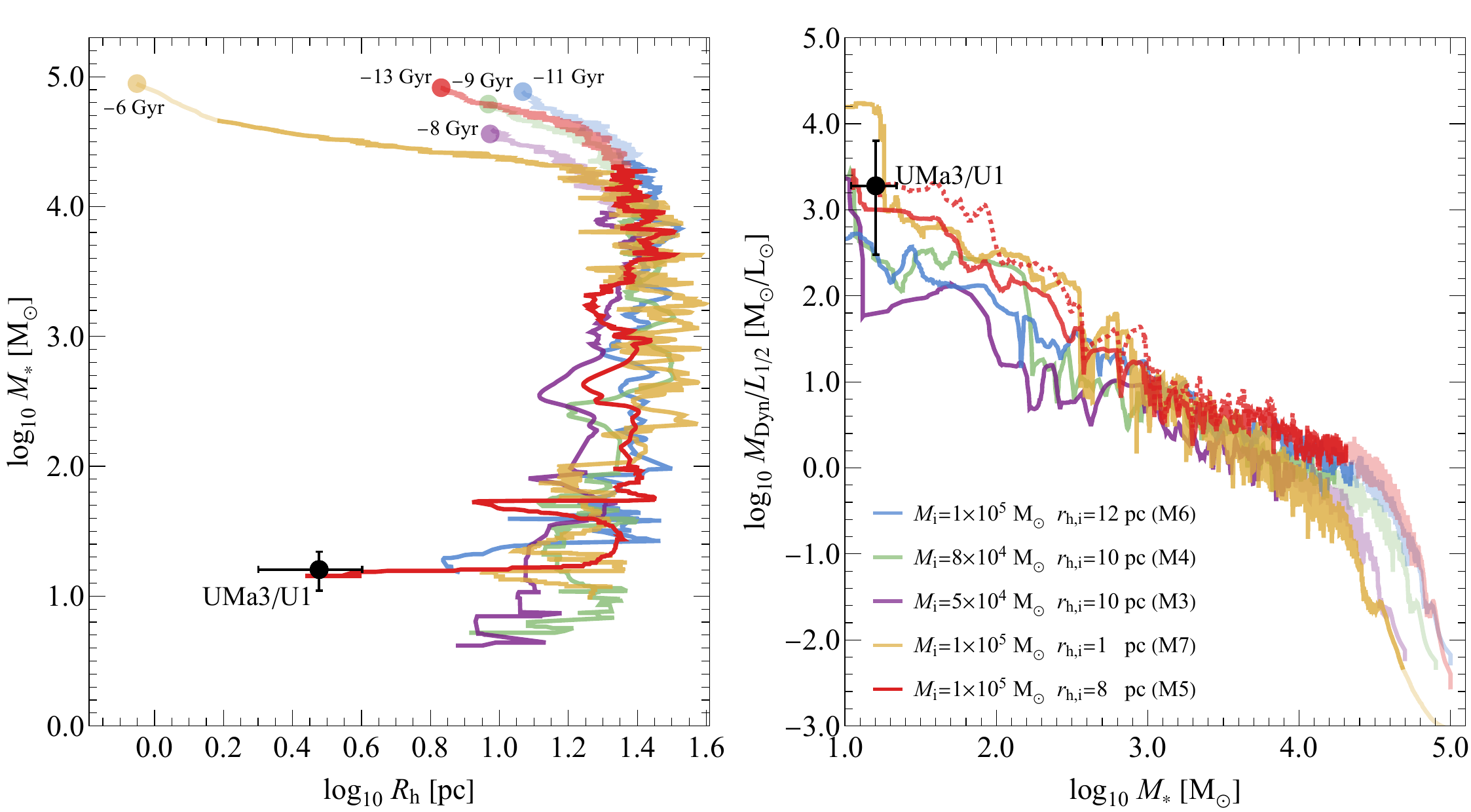}  
    \caption{Evolutionary tracks of five simulated clusters, benchmarked against UMa3/U1. \textit{Left:} Cluster trajectories in the $M_*$–$\Rh$ plane. \textit{Right:} Progression of $M_\mathrm{Dyn}/L_{1/2}$ with stellar mass loss. Dynamical masses are computed from bound stellar sums within $\Rh$; for model M5, an additional estimate using Equation~\ref{eq:mass} based on the mock-observed velocity dispersion from the UNIONS range is shown as the red dashed line. Statistical smoothing is applied in both panels to reduce snapshot-to-snapshot fluctuations. Faint (solid) segments indicate evolutionary stages before (after) entering the DSC phase.} 
    \label{fig:UMa3}
\end{figure*}

\begin{figure}
  \centering
  \includegraphics[scale=0.6]{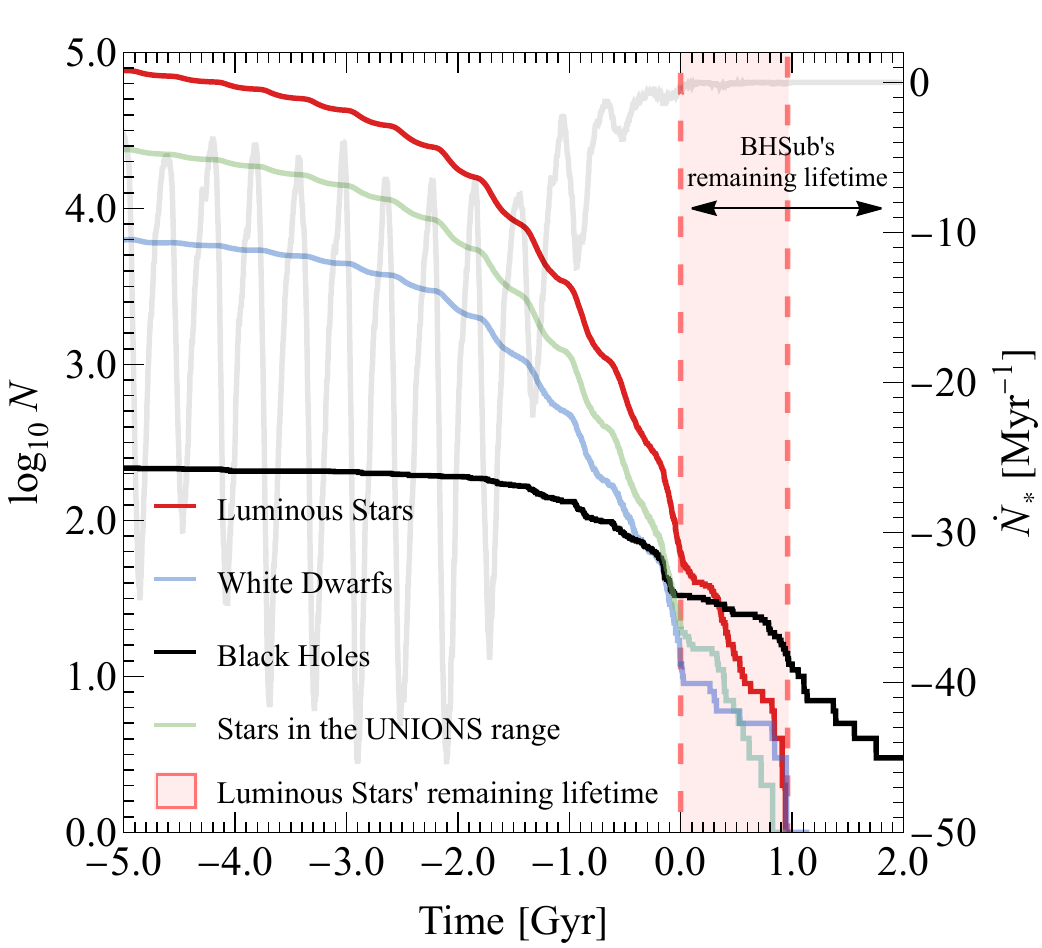}
  \caption{Temporal evolution of the M5 model’s stellar content: luminous stars (red), subset within the UNIONS detection limits (green), white dwarfs (blue), and black holes (black). The shaded pink band marks the remaining lifetime of bound luminous stars. The gray curve (right axis) represents the instantaneous escape rate of luminous stars ($\overset{\cdot}{N}_*$), which drops sharply after $12.5 \Gyr$ as the remaining stars become deeply bound within the BHSub.}
  \label{fig:Numbers}
\end{figure}

Our results are divided into two subsections. First, we compare the simulated clusters with the observed parameters of UMa3/U1 to assess whether the DSC phase can simultaneously account for its high $M_\mathrm{Dyn}/L$ and compactness (\secref{sec:UMa3}). We then broaden our analysis, exploring the $\Rh$–$L$ and $M_\mathrm{Dyn}/L$–$L$ parameter spaces to delineate DSC-occupied regions and evaluate their capacity to encompass faint, ambiguous satellites (\secref{sec:General}).

\subsection{UMa3/U1}\label{sec:UMa3}

The DSC phase emerges from the interplay between BHSub self-depletion and luminous star evaporation timescales. Thus, the initial cluster parameters that influence these timescales dictate the transition. In our simulations, mimicking UMa3/U1's orbit and metallicity, the $r_\mathrm{h,i}$ and IMF (specifically, BH retention fraction) play a key role in reaching this phase. With a canonical IMF, a higher initial density makes DSC phase entry less probable \citep{Gieles2023,DSC}. A deeper gravitational potential traps low-mass luminous stars, preventing the energy imparted by the BHSub from propelling them to escape velocity, thereby limiting their evaporation rate. The size of a cluster’s BH population depends on the high-mass end of the stellar IMF. A top-heavy IMF produces a more massive BHSub, enhancing dynamical heating through more frequent few-body encounters and injecting additional energy into the background stars, triggering the evolution to the super-virial phase even in dense environments \citep{Breen2013,Longwang2020,DSC}. Consequently, the canonical IMF models were set as low density to accelerate the dominance of BHSub over luminous stars and increase $M_\mathrm{Dyn}/L$, while ensuring that $r_\mathrm{h,i}$ remained constrained to match the final compactness of UMa3/U1. Conversely, top-heavy IMF models were initialized with a small  $r_\mathrm{h,i}$ to counteract their substantial expansion driven by high energy injection from the BHSub.

To compare the modeled clusters with UMa3/U1, we analyze their final snapshots at $M_*=16\Msun$. We then evaluate their $\tau$, $\Rh$, and $M_\mathrm{Dyn}/L_{1/2}$, verifying whether they meet the thresholds $\tau \gtrsim 11\Gyr$, $\Rh=3\pm1\pc$, and $M_\mathrm{Dyn}/L_{1/2}=1900^{+4400}_{-1600} \ \mathrm{M_{\odot}/L_{\odot}}$. The results are presented in \tabref{tab:initial_conditions}. The left panel of \figref{fig:UMa3} illustrates the evolutionary trajectories of the five modeled clusters in the mass-size plane. BHSub-driven energy injection expands the canonical and top-heavy IMF clusters to 30 and 40 pc, respectively. In the final stages, as a substantial fraction of luminous stars evaporates and only a few remain bound around the BHSub, $\Rh$ contracts to less than 10 pc. The right panel of \figref{fig:UMa3} depicts the variation of the $M_\mathrm{Dyn}/L_{1/2}$ as a function of stellar mass loss. As the clusters evolve, $M_\mathrm{Dyn}/L_{1/2}$ increases steadily, exceeding 2500 in the final stages. During the DSC phase, the high evaporation rate of luminous stars gradually increases the dark remnant fraction of the cluster. As the dark population becomes dominant, the $M_\mathrm{Dyn}/L_{1/2}$ rises dramatically, climbing above $10^4$ in top-heavy models.

\figref{fig:UMa3} reveals that while all modeled clusters reach the $M_\mathrm{Dyn}/L_{1/2}$ threshold of UMa3/U1, the majority fail to fall within its occupied region in the $\Rh$–$M_*$ plane. In canonical IMF models, the initial cluster density is insufficient for significant late-stage contraction. In contrast, although the top-heavy models start out dense, the BHSub-induced heating makes it challenging for them to remain compact in their final stages. Optimizing initial parameters, such as $r_\mathrm{h,i}$, $M_\mathrm{i}$, BH natal kicks, and IMF, enables the formation of various DSC progenitors that also align with UMa3/U1 in terms of $\Rh$. However, the heavy computational burden of \Nbody collisional simulations makes systematic optimization highly demanding. Among our modeled clusters, M5 (\figref{fig:UMa3}: red lines) more accurately reproduces the observed parameters of UMa3/U1, making it a strong progenitor candidate. This model indicates that DSCs can provide a formation scenario for UMa3/U1 without invoking dark matter, reinforcing its classification as a self-gravitating star cluster. In fact, the initial values for this model, M5, are consistent with the post-gas-expulsion re-virialized half-mass radii of star clusters born as embedded clusters with a canonical IMF \citep{Haghi2015, Dinnbier2020, Wu2018, Wu2019}, making M5 particularly realistic.

The cluster entered the DSC phase $\approx 4 \Gyr$ ago ($\widetilde{\mathit{\tau}}_\mathrm{DSC}=0.35$) and will dissolve entirely in the next $2 \Gyr$, as its BHSub undergoes disruption. \figref{fig:Numbers} illustrates the temporal evolution of the cluster’s stellar content. Once the bound stellar mass declines to $M_* = 16\Msun$, 62 luminous stars remain gravitationally bound and are expected to fully escape the system over the following about $ 1\Gyr$. The sharp decline in the stellar escape rate beyond $12.5\Gyr$ results from the depletion of loosely bound stars, whose earlier evaporation is driven by energy injected from the central BHSub. The remaining luminous stars are tightly bound within the deep gravitational potential of the BHSub and therefore require significantly longer timescales to escape, as reflected in the flattening of the gray curve in \figref{fig:Numbers}. Also shown in the same figure is the evolution of stars within the detection limits of the UNIONS survey, defined by an $i$-band apparent magnitude range of 17.5–23.5 mag (corresponding to stellar masses between 0.31 and 0.79 $\Msun$ \citet{Devlin2025}). These stars, referred to as the 'UNIONS range', are depleted on a slightly shorter timescale of about $820$ Myr.

\citet{Errani2024} derived a line-of-sight velocity dispersion of $\sigma_\mathrm{los}=49 \ \mathrm{m \ s}^{-1}$ for UMa3/U1 using the projected virial theorem \citep{Errani2018}, based on its measured $M_*$ and $\Rh$, thereby ruling out the possibility that this system is a self-gravitating star cluster. Likewise, \citet{Smith2024} obtained a comparable estimate via the \citet{Wolf2010} mass estimator:
\begin{equation}
M_{1/2} \approx 930 \left(\frac{\sigma_\mathrm{los}^2}{\mathrm{km^2\,s^{-2}}}\right) \left(\frac{R_\mathrm{h}}{\mathrm{pc}}\right) \ \mathrm{M_\odot},
\label{eq:mass}
\end{equation}
where $M_{1/2}$ is the dynamical mass enclosed within $\Rh$. However, in clusters where the core is dominated by a BHSub, velocity dispersion predictions based on the surface mass density profile fail to capture the actual dynamics and tend to be systematically underestimated \citep{Wang2016,Wu2024}. In other words, the underestimation of predicted velocity dispersion suggests the presence of a hidden mass component in the core. Our results indicate that, assuming UMa3/U1 is a self-gravitating DSC, its line-of-sight velocity dispersion could naturally reach the lower bound of observed estimates, corresponding to $M_\mathrm{Dyn}/L_{1/2}\approx10^3 \ \mathrm{M_{\odot}/L_{\odot}}$. Moreover, \citet{Errani2024} did not include stellar evolution in their self-gravitating models, which likely accounts for their prediction of only a single orbital passage before complete disruption. In contrast, our models incorporate a BHSub, which enhances the central mass concentration, deepens the gravitational potential, and increases the overall binding energy, consequently extending the system's survival time to nearly 2.5 times longer than the disruption timescale predicted by \citet{Errani2024}.

We note that the values of $M_\mathrm{Dyn}/L_{1/2}$ shown in the left panel of \figref{fig:UMa3} are computed by summing the masses of all bound stars within the $\Rh$. For comparison, we also mock-observe the dynamical mass of model M5 using Equation~\ref{eq:mass}. Stellar velocities are measured relative to the cluster’s center of mass within the UNIONS detection limits; the three velocity components are merged into a single one-dimensional sample, and the standard deviation is taken to estimate the $\sigma_\mathrm{los}$. The resulting $M_\mathrm{Dyn}/L_{1/2}$ (red dashed line in \figref{fig:UMa3}) closely matches the simulation-based value for $M_* \gtrsim 10^3~\Msun$, though it slightly overestimates it at lower $M_*$.

\subsection{Mapping DSCs in \texorpdfstring{$\Rh$–$L$ and $M_\mathrm{Dyn}/L$–$L$}{Rh-L and M_Dyn/L-L} parameter spaces}\label{sec:General}

In this section, we broaden our analysis by tracing the evolutionary tracks of DSCs within structural parameter spaces, assessing whether they can naturally descend into regions typically occupied by faint, ambiguous satellites, and thereby constitute a plausible formation scenario for such objects. We explored the evolution of a model transitioning into the DSC phase using the \textsc{NBODY7} code. The model adopts a canonical IMF and is initialized with $M_\mathrm{i} = 6\times10^4 \ \mathrm{M_{\odot}}$ and $r_\mathrm{h,i} = 5 \ \mathrm{pc}$. A metal-poor composition is assumed to enhance the BH retention fraction \citep{Belczynski2008,Belczynski2010}. The cluster is placed at Galactocentric distances of $R_\mathrm{G} \ (\kpc) \in\{8,16\}$ on circular orbits, with the remaining setup parameters taken from \secref{sec:sim}. We emphasize that clusters spanning a wide array of initial conditions are capable of evolving into the DSC phase, resulting in DSCs occupying a much broader region in parameter spaces compared to the narrower range covered by our models.

\begin{figure*}
    \centering
    \includegraphics[width=1\linewidth]{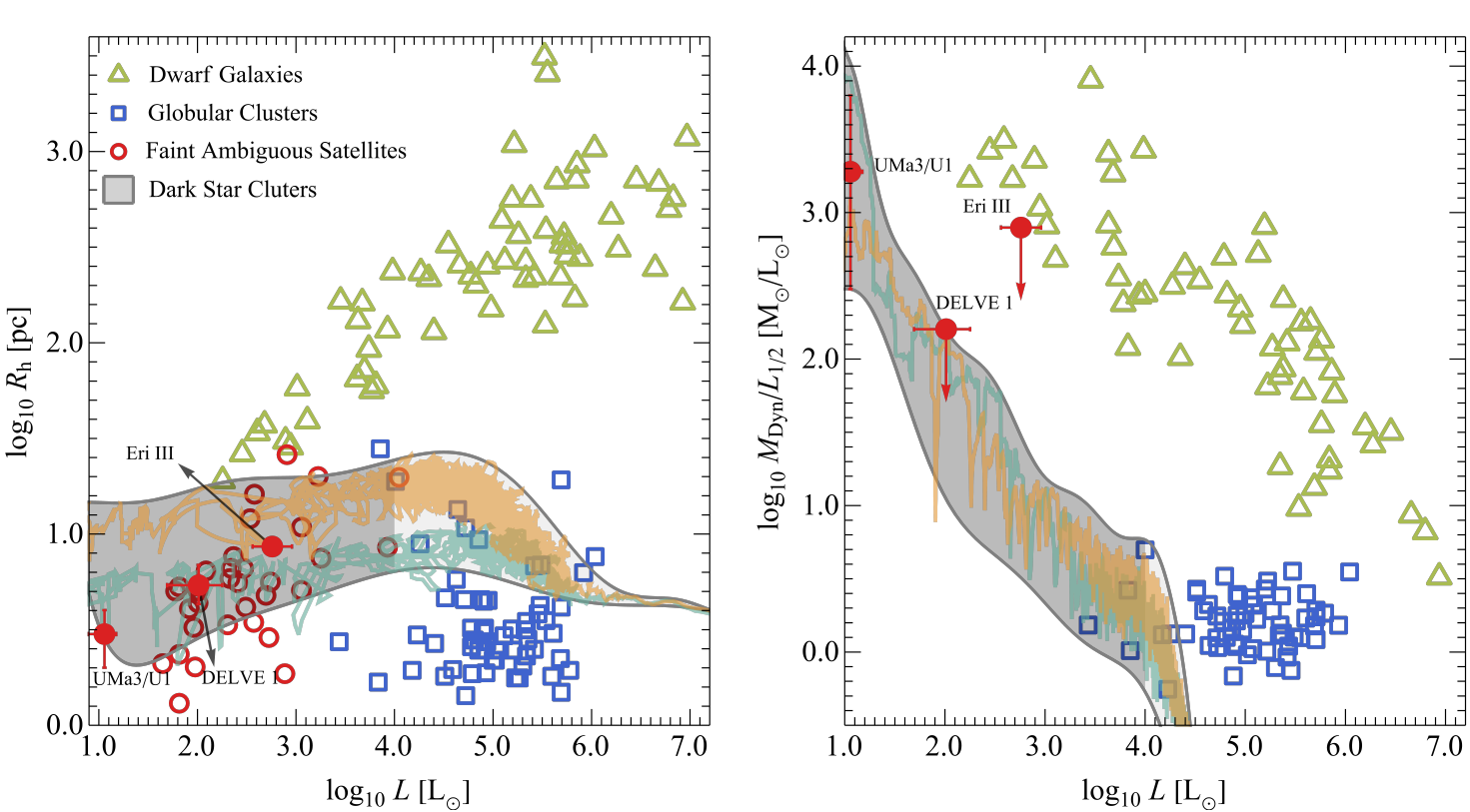}  
    \caption{Compilation of Local Group dGs (\citealt{McConnachie2012}; version 2021 January), GCs (\citealt{Harris1996}; version 2010 December), and faint, ambiguous satellites \citep{Cerny2023ref,Cerny2023,Smith2024,Simon2024} in the $\Rh$–$L$ (left) and $M_\mathrm{Dyn}/L_{1/2}$–$L$ (right) parameter spaces. Solid lines trace the tidal evolutionary tracks of DSC models (Sec. 3.2) on circular orbits at $R_\mathrm{G} = 8\kpc$ (green) and $R_\mathrm{G} = 16\kpc$ (orange). The gray shaded region spans the full parameter space swept by the DSC models along their evolutionary tracks. The darker portion marks the period during which the clusters are in the DSC phase ($Q_* \geq 1$). Dynamical mass is obtained by summing bound stellar plus remnant masses within $\Rh$ in both models. Filled symbols indicate observed ambiguous satellites with velocity dispersion measurements.} 
    \label{fig:General}
\end{figure*}

\figref{fig:General} (left panel) displays a compilation of Local Group dGs and GCs in the $\Rh$–$L$ diagram. Despite the distinct distribution of these satellite systems, the separation breaks down at low total luminosities, where faint stellar systems occupy an overlapping, ambiguous region. The simulated clusters initially pass through the GC's region and progressively enter the ambiguous zone within the faint regime. The gray-shaded band indicates the extent covered by the models along their evolutionary tracks, encompassing the faint, ambiguous satellites. Due to the smaller tidal radius of the simulated cluster closer to the Galactic center, its expansion is restricted, establishing the lower limit of the band. Varying the DSC models' initial parameters can expand the band to span $\Rh \approx 1-50 \pc$  at low total luminosities.

\begin{figure}
  \centering
  \includegraphics[scale=0.6]{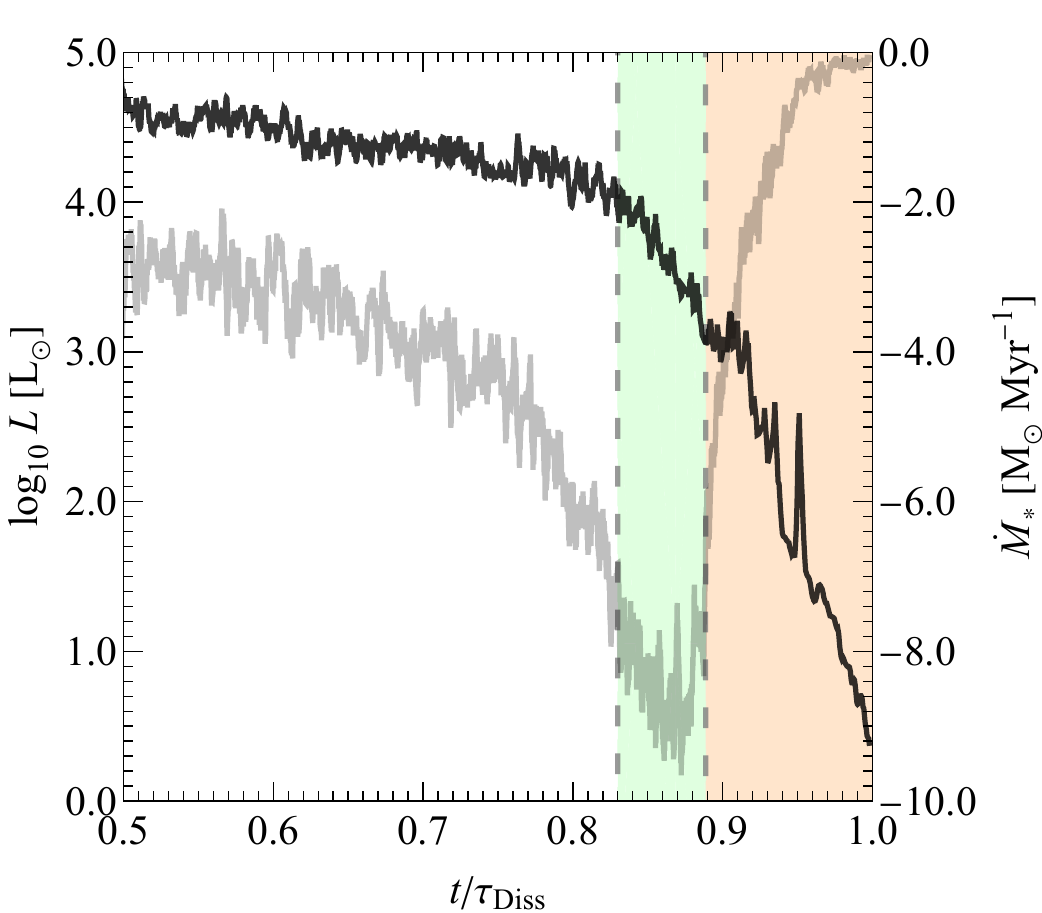}
  \caption{
Time-resolved evolution of the model shown in \figref{fig:General} (orange line), located at $R_\mathrm{G} = 16\kpc$. The black curve (left axis) shows the cluster’s luminosity as a function of normalized evolutionary time ($t/\tau_\mathrm{Diss}$), indicating the fraction of its lifetime spent at each luminosity level. The gray curve (right axis) shows the instantaneous mass-loss rate of luminous stars ($\overset{\cdot}{M}_*$) throughout the cluster’s evolution. The green shaded band highlights the transitional luminosity regime ($10^3 < L/\mathrm{L}_\odot < 10^4$), where the cluster evolves rapidly, while the orange band marks the faint regime ($L < 10^3~\mathrm{L}_\odot$), where evolution proceeds more slowly.}
  \label{fig:TimeEvolution}
\end{figure}

In the right panel of \figref{fig:General}, satellites are plotted in the $M_\mathrm{Dyn}/L_{1/2}$ versus $L$ space. To estimate their dynamical mass enclosed within $\Rh$, we employed the \citet{Wolf2010} mass estimator (Equation~\ref{eq:mass}). GCs follow a tight sequence along $M_\mathrm{Dyn}/L_{1/2}\approx 2 \ \mathrm{M_{\odot}/L_{\odot}}$, whereas dGs are offset to significantly higher values, scattering across $M_\mathrm{Dyn}/L_{1/2} \approx 10-10^4 \ \mathrm{M_{\odot}/L_{\odot}}$. Owing to the lack of follow-up spectroscopy for most faint, ambiguous satellites, the velocity dispersion has been calculated for only a handful of objects. For the case of UMa3/U1, taking $\sigma_\mathrm{los} = 1.9 \ \kms$ at face value (removing the furthest velocity outlier, as in \citealt{Smith2024}) results in an $M_\mathrm{Dyn}/L_{1/2} = 1900 \ \mathrm{M_{\odot}/L_{\odot}}$. Eridanus III and DELVE 1 have also been analyzed; however, large kinematic uncertainties allowed only weak upper limits on $\sigma_\mathrm{los}$ \citep{Simon2024}: $<5.4$ (1.2) $\kms$, corresponding to $M_\mathrm{Dyn}/L_{1/2} < 790 $ ($160$) $\mathrm{M_{\odot}/L_{\odot}}$ for Eridanus III (DELVE 1). As depicted in the right panel of \figref{fig:General}, these ambiguous satellites lie within the low-luminosity tail of the dark matter–dominated sequence.

Applying the modeled cluster outlines in the $M_\mathrm{Dyn}/L_{1/2}-L$ diagram (\figref{fig:General}, right panel: gray-shaded band) reveals that DSCs trace a transitional channel bridging GCs and dGs, passing through the accumulation of GCs ($M_\mathrm{Dyn}/L_{1/2}\approx 2 \ \mathrm{M_{\odot}/L_{\odot}}$) and rising sharply to $M_\mathrm{Dyn}/L_{1/2} \approx 10^4 \ \mathrm{M_{\odot}/L_{\odot}}$ as BHs dominate. The $M_\mathrm{Dyn}/L_{1/2}$ evolution of both models is initially comparable; however, the cluster at $R_\mathrm{G} = 8\kpc$ (green line) ultimately exhibits higher $M_\mathrm{Dyn}/L_{1/2}$. This is attributed to stronger tidal stripping near the Galactic center, which accelerates the evaporation of low-mass luminous stars and shifts the mass budget toward BH remnants \citep{DSC}. The gray band in the right panel of \figref{fig:General} highlights the DSC channel’s capacity to capture faint, ambiguous satellites with elevated $M_\mathrm{Dyn}/L$. Expanding the model set, especially by including those adopting a top-heavy IMF, would broaden this channel, filling a greater portion of the high-$M_\mathrm{Dyn}/L$ space in the faint regime.

Our results suggest that DSCs can account for the nature of faint, ambiguous satellites with dG–like $M_\mathrm{Dyn}/L$, without invoking dark matter. In this scenario, the progenitors originate in the GC region of parameter space, contrasting with the microgalaxy perspective, in which the evolutionary path begins in the dG domain and ultimately engulfs the ambiguous population \citep{Errani2024}. This result is important also in view of the Chandrasekhar dynamical friction test \citep{Roshan2021, OehmKroupa2024}.

The apparent gap in the $\Rh$–$L$ diagram between GCs and ambiguous satellites, particularly in the luminosity range $10^3 \lesssim L/\mathrm{L}_\odot \lesssim 10^{4}$, may be a natural consequence of the rapid dynamical evolution of DSCs through this regime. \figref{fig:TimeEvolution} shows the time-resolved evolution of the model at $R_\mathrm{G} = 16\kpc$; luminous stars are rapidly ejected from the cluster once it enters the super-virial phase, leading to a short-lived evolutionary stage characterized by enhanced stellar mass loss. During this interval, the cluster’s luminosity drops sharply as luminous stars escape at the highest rate recorded throughout its evolution. This is quantified in \figref{fig:TimeEvolution} (right axis), where the luminous stellar mass-loss rate $\overset{\cdot}{M}_*$ peaks precisely within the $10^3$–$10^{4}~\mathrm{L}_\odot$ range. Despite spanning a wide range in luminosity, this episode accounts for only $\approx5.5 ~\%$ of the cluster’s lifetime. As loosely bound luminous stars are stripped away, the surviving population becomes tightly bound within the BHSub’s potential well, causing the escape rate to drop significantly. This shift extends the duration the cluster spends in the faint regime ($L \lesssim 10^3~\mathrm{L}_\odot$) to $\approx11.5~\%$ of its lifetime. In top-heavy IMF models, the residence time in both luminosity regimes increases. For example, model M7 ($\alpha_3 = 1.7$) spends $\approx40 ~\%$ of its lifetime in the faint regime, and less than half of that in the $10^3$–$10^{4}~\mathrm{L}_\odot$ range. This time-asymmetry between the rapid transition through the intermediate-luminosity regime and the slower evolution in the faint regime provides a natural explanation for the observed scarcity of systems in the luminosity gap.

It should be mentioned that the ratio of the time that a cluster spends with a luminosity in the range $10^3$–$10^{4}~\mathrm{L}_\odot$ to the time it spends below a luminosity of $L \lesssim 10^3~\mathrm{L}_\odot$ is approximately $1:2$. In comparison,  \figref{fig:General} shows that the number of observed clusters within the $10^3$–$10^{4}~\mathrm{L}_\odot$ luminosity range relative to those with luminosities below $L \lesssim 10^3~\mathrm{L}_\odot$ is roughly $1:3$. This observation is consistent with the expected time that a cluster spends in each luminosity range.

\section{Summary and Conclusions}\label{sec:conclu}

The recently discovered UMa3/U1 stands out as one of the rare faint, ambiguous satellites with a measured spectroscopic velocity dispersion, shedding light on the nature of such systems. Although its size is comparable to that of most GCs ($\Rh=3\pm1\pc$), its inferred dynamical mass-to-light ratio is unusually high ($M_\mathrm{Dyn}/L_{1/2}=1900^{+4400}_{-1600} \ \mathrm{M_{\odot}/L_{\odot}}$), characteristic of apparent dark matter-dominated dGs. High $M_\mathrm{Dyn}/L$, typically associated with galaxies, can also arise in DSCs, where a centrally segregated BHSub that is retained due to low natal kicks, heats the stellar population and drives the loss of luminous stars producing BH-dominated and thus a super-virial appearance and mimicking a high $M_\mathrm{Dyn}/L$.

To assess whether self-gravitating star clusters evolving into the DSC phase could be the progenitors of UMa3/U1, replicating both its high $M_\mathrm{Dyn}/L_{1/2}$ and compactness, we conducted direct \Nbody simulations calibrated to UMa3/U1’s orbit and metallicity. Despite reaching the target $M_\mathrm{Dyn}/L_{1/2}$ threshold, most modeled clusters fail to match UMa3/U1’s compactness, primarily due to expansion driven by BHSub energy injection. Among our simulations, the model with $M_\mathrm{i} = 10^5\Msun$ and  $r_{\mathrm{h,i}} = 8\pc$ most accurately reproduces UMa3/U1 (\figref{fig:UMa3}), positioning it as a strong progenitor candidate. This scenario eliminates the need for dark matter and reinforces UMa3/U1’s identity as a self-gravitating star cluster. Optimizing the initial setup can generate multiple DSC progenitors that also align with UMa3/U1’s $\Rh$, though at high computational cost.

We proceed by mapping the evolutionary trajectories of DSC models in the $\Rh$–$L$ and $M_\mathrm{Dyn}/L_{1/2}$–$L$ parameter spaces, outlining DSC-occupied regions and evaluating their overlap with faint, ambiguous satellite systems. In the $\Rh$–$L$ plane, DSCs initially traverse the GCs' region and ultimately sweep across the ambiguous zone within the faint regime (left panel of \figref{fig:General}). DSCs trace a transitional channel bridging GCs and dGs in the $M_\mathrm{Dyn}/L_{1/2}$–$L$ diagram as they rise from $M_\mathrm{Dyn}/L_{1/2} \approx 2$ to $10^4 \ \mathrm{M_{\odot}/L_{\odot}}$ (right panel of \figref{fig:General}), highlighting the DSC channel’s capacity to encompass faint, ambiguous satellites with elevated $M_\mathrm{Dyn}/L$.

In conclusion, this paper proposes a purely baryonic origin for faint, ambiguous satellites with high  $M_\mathrm{Dyn}/L$, suggesting they evolve from GCs with long-lived BHSub offering an alternative to the dark matter–dominated microgalaxy scenario. Interpreting faint, ambiguous satellites as DSCs carries several broader implications. It suggests that some star clusters may have formed with a top-heavy IMF \citep{MarksMichael2012}, which promotes the early formation of a substantial BH population and facilitates evolution into the DSC phase with elevated $M_\mathrm{Dyn}/L$. It also implies low natal kicks for BHs, leading to their long-term retention and central concentration into a BHSub. This, in turn, lends further support to a dynamical cluster-origin pathway for BH mergers (e.g., \citealt{Wang2021,Weatherford2021}) and the BH-driven formation of dense stellar streams (e.g., \citealt{Gieles2021}).

\section*{Data availability}
The data underlying this paper are available in the paper.
\begin{acknowledgments}
This work is based upon research funded by Iran National Science Foundation (INSF) under project No.4035689. AR acknowledges support from the Iran National Science Foundation. HH thanks the Argelander Institute for Astronomy (AIfA) and the SPYDOR group for their hospitality. AHZ acknowledges support from the Alexander von Humboldt Foundation. PK thanks the DAAD Bonn-Prague Eastern European Exchange Program.  
\end{acknowledgments}


\bibliography{sample7}{}
\bibliographystyle{aasjournalv7}



\end{document}